\documentclass[aps,twocolumn,showpacs,preprintnumbers,amsmath,amssymb,superscriptaddress]{revtex4}

\usepackage{graphicx}
\usepackage{dcolumn}
\usepackage{bm}
\usepackage{hyperref}
\usepackage{latexsym}

\begin{document}

\title{Power-law temporal auto-correlations in day-long records of human
physical activity and their alteration with disease}

\author{Lu\'{\i}s A. Nunes \surname{Amaral}}
\email{amaral@northwestern.edu}
\homepage{http://amaral.chem-eng.northwestern.edu/}
\affiliation{Dept. of Chemical Engineering, Northwestern University, 
             Evanston, IL 60208}
\affiliation{Center for Polymer Studies and Dept. of Physics,
             Boston University, Boston, MA 02215}

\author{Danyel J. Bezerra Soares}
\affiliation{International Center for Complex Systems and 
	     Dept. F\'{\i}sica Te\'orica e Experimental,
	     Universidade Federal do Rio Grande do Norte, Natal, Brazil}

\author{Luciano R. da Silva}
\affiliation{Center for Polymer Studies and Dept. of Physics,
             Boston University, Boston, MA 02215}
\affiliation{International Center for Complex Systems and 
	     Dept. F\'{\i}sica Te\'orica e Experimental,
	     Universidade Federal do Rio Grande do Norte, Natal, Brazil}

\author{Liacir S. Lucena}
\affiliation{International Center for Complex Systems and 
	     Dept. F\'{\i}sica Te\'orica e Experimental,
	     Universidade Federal do Rio Grande do Norte, Natal, Brazil}

\author{Mariko Saito}
\author{Hiroaki Kumano}
\affiliation{Dept. of Psychosomatic Medicine, The University of Tokyo
Hospital, Tokyo 113-8655, Japan}

\author{Naoko Aoyagi} 
\affiliation{Educational Physiology Laboratory, Graduate School of Education,
The University of Tokyo, Tokyo 113-0033, Japan}

\author{Yoshiharu Yamamoto} 
\affiliation{Educational Physiology Laboratory, Graduate School of Education,
The University of Tokyo, Tokyo 113-0033, Japan}
\affiliation{PRESTO, Japan Science and Technology Corporation, Saitama
332-0012, Japan}

\begin{abstract}

We investigate long-duration time series of human physical activity under
three different conditions: healthy individuals in (i) a constant routine
protocol and (ii) in regular daily routine, and (iii) individuals diagnosed
with multiple chemical sensitivities.  We find that in all cases human
physical activity displays power law decaying temporal auto-correlations.
Moreover, we find that under regular daily routine, time correlations of
physical activity are significantly different during diurnal and nocturnal
periods but that no difference exists under constant routine
conditions. Finally, we find significantly different auto-correlations for
diurnal records of patients with multiple chemical sensitivities.

\end{abstract}


\maketitle


Healthy free-running physiologic systems have complex self-regulating
mechanisms which process inputs with a broad range of characteristics
\cite{Shlesinger}, and may generate signals that have scale-invariant
dynamics \cite{Musha,Peng95}.  Physical activity is a physiologic signal of
great interest due to its (i) impact on other rhythms such as heart rate, and
(ii) significance for certain psychiatric or psychosomatic conditions
\cite{Teicher95,psy}.  Indeed, a number of researchers has used actigraph
recordings to classify types of motion---such as sitting, moving or standing
\cite{children}---with the goal of developing ambulatory methods for the
diagnosing of attention-deficit hyperactivity disorder in children
\cite{adhd}.  In such applications, the advantage of using actigraph
recordings is that it may replace more limiting, cumbersome, and intrusive
techniques such as videotaping \cite{monitoring}.  Surprisingly, up to now
there has been no attempt to quantify the time correlations of the
fluctuations in the levels of physical activity in ``free-running''
ambulatory settings or their alteration under special conditions, such as
constrained activity or psychiatric disease \cite{Teicher}.

In this Letter, we study long-duration time series of human physical activity
under two different conditions: a constant routine (CR) protocol where
physical activity and postural changes are kept to a minimum
\cite{Constant_rout}, and (ii) regular daily routine (RDR).  We find that
physical activity displays power law decaying temporal correlations for these
conditions.  Moreover, we find that during diurnal periods the time
correlations of physical activity are significantly different for regular
daily routine and CR protocols.  In contrast, we find no differences between
diurnal and nocturnal periods for the CR protocol while we find significant
differences between diurnal and nocturnal periods for the regular daily
routine.  Further, we find similarities between the CR patterns of human
physical activity and those of nocturnal periods for regular daily
routine \cite{note-cr}.

Our finding of long-range time-correlations in activity of healthy
individuals under regular daily activity and the change in those correlation
properties under controlled conditions, suggests that possibility that the
fractal index characterizing the long-range auto-correlations of the time
series of physical activity may be an additional probe \cite{Teicher95} of
the abnormal patterns of physical activity of patients with psychiatric
disorders such as depression, bipolar disorder, or schizophrenia \cite{DSM},
and with psychosomatic illnesses associated with exaggerated fatigue.  We
test this possibility with data collected for individuals diagnosed with
multiple chemical sensitivities (MCS).  This is a disease \cite{mcs_criteria}
characterized by a positive chemical exposure history and multi-system
symptoms, such as fatigue, headache, sleep disturbance, and myalgia, which
are elicited after exposure to various chemical compounds
\cite{buchwald94}. The clinical picture of MCS is shown \cite{buchwald94} to
largely overlap with that of other psychosomatic illnesses with severe
fatigue symptoms \cite{holmes88,fukuda94}.  Interestingly, we find a {\it
significant difference\/} in the degree of diurnal auto-correlations in
physical activity for MCS patients and healthy controls leading a regular
daily routine.

%
\begin{figure}[t!]
 \vspace*{-0.5cm} 
 \includegraphics*[width=7.6cm]{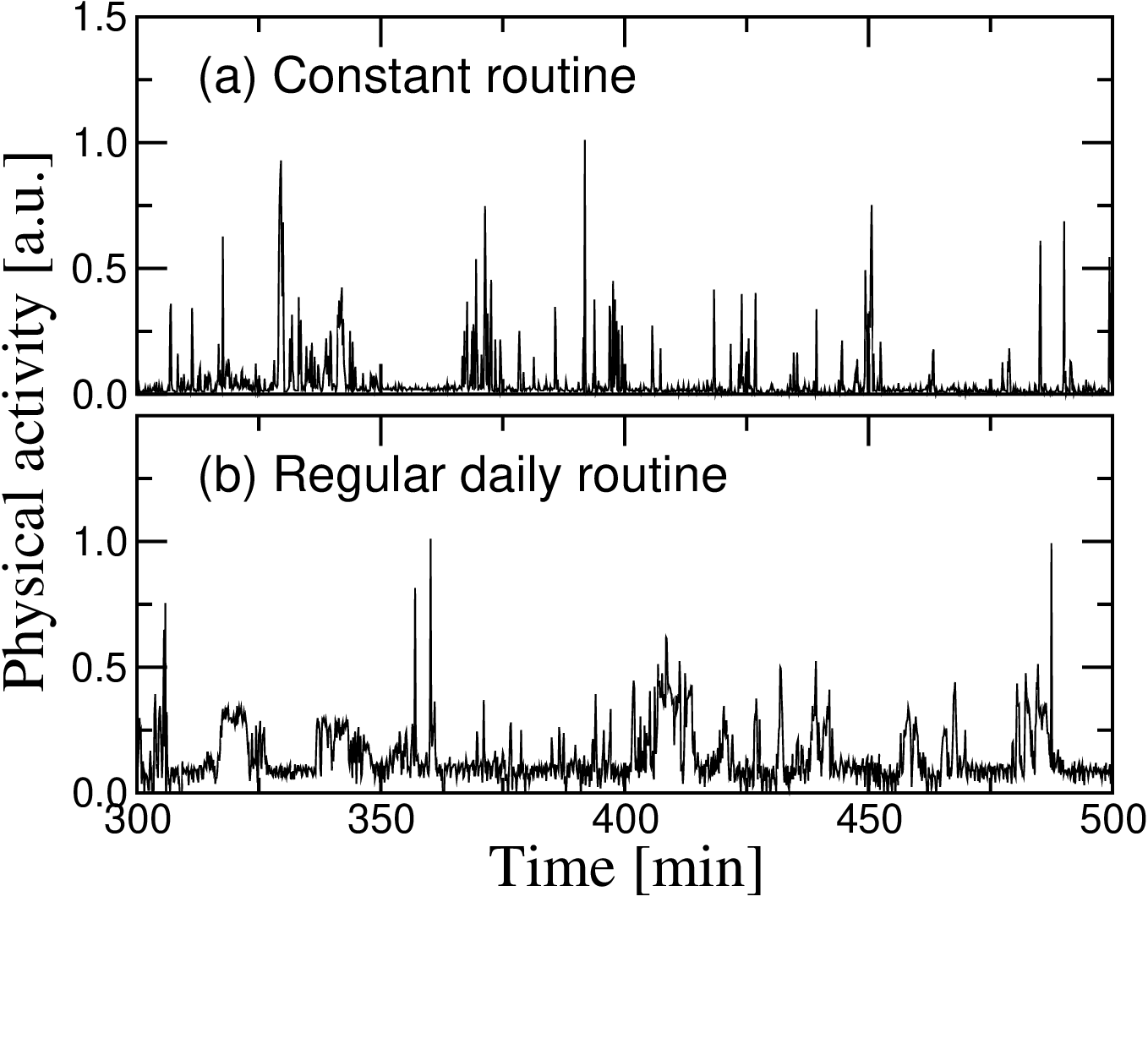} 
 \vspace*{-0.5cm} 
 \caption{ Magnitude of physical activity in arbitrary units (a.u.)  for (a)
  subject \#2 during CR conditions, and (b) the same subject during regular
  daily routine.  Note how the episodes of ``sustained'' physical activity
  are reduced during CR conditions.  }
 \label{f.signals}
\end{figure}
%


We analyze datasets from seven healthy, non-smoking male subjects
\cite{Tokyo} (ages: 21--30~yr).  For six of the subjects, we obtained
two datasets per subject, the first under CR conditions, and the
second under regular daily routine conditions
\cite{Constant_rout,na00,Amaral01}.  We recorded \cite{na00} body
acceleration using a portable, {\it waist-worn}, long-term ambulatory
monitoring device (Amx720, Nihon-Koden Wellness Corporation, Japan).
The monitoring device has two shock sensors that measure body
acceleration at the waist in the vertical and horizontal directions
with a sensitivity of 0.08$g$.  The acceleration was
recorded after being full-wave rectified and integrated over 8~s
intervals \cite{na00}. In the present study, we consider only body
acceleration in the vertical direction.

Figure~\ref{f.signals} displays vertical acceleration at the waist for
subject \#2 for both CR protocol and regular daily routine, Vertical
acceleration at the waist is a good proxy for physical activity
because it correlates to {\it entire body motion\/}.
The figure suggests a decrease in the number of episodes of sustained
physical activity during CR but the quantitative nature of such time
organization is not clear from the graph. Thus, we apply the detrended
fluctuations analysis (DFA) method to quantify long-range
time-correlations in the physical activity time series
\cite{Peng95,Ashkenazy01}.  The DFA method is defined as follows: One
first integrates the physical activity time series.  One then divides
the time series into ``boxes'' of length $n$ and perform, in each box,
a least-squares polynomial fit of order $k$ fit to the integrated
signal.  Next, one calculates in each box the root-mean-square
deviations of the integrated signal from the linear fit.  The basic
idea is that the polynomial fit represents the local trend in each box
while the magnitude of the fluctuations around the trend signals the
degree of the time correlations in the signal.

This procedure is repeated for different box sizes (time scales) $n$.  For
fractal signals one finds a power-law relation between the average magnitude
of the fluctuations $F(n)$ and the number of points $n$
\begin{equation}
F(n) \sim n^{\alpha} \,,
\label{e.dfa}
\end{equation}
where the scaling exponent $\alpha$ quantifies the degree of the
correlations.  Uncorrelated time series yield $\alpha = 0.5$, long-range
anti-correlations result in $\alpha < 0.5$, and long-range positive
correlations results in $\alpha > 0.5$.

\begin{figure}[t!]
 \vspace*{0.0cm}
 \includegraphics*[width=7.6cm]{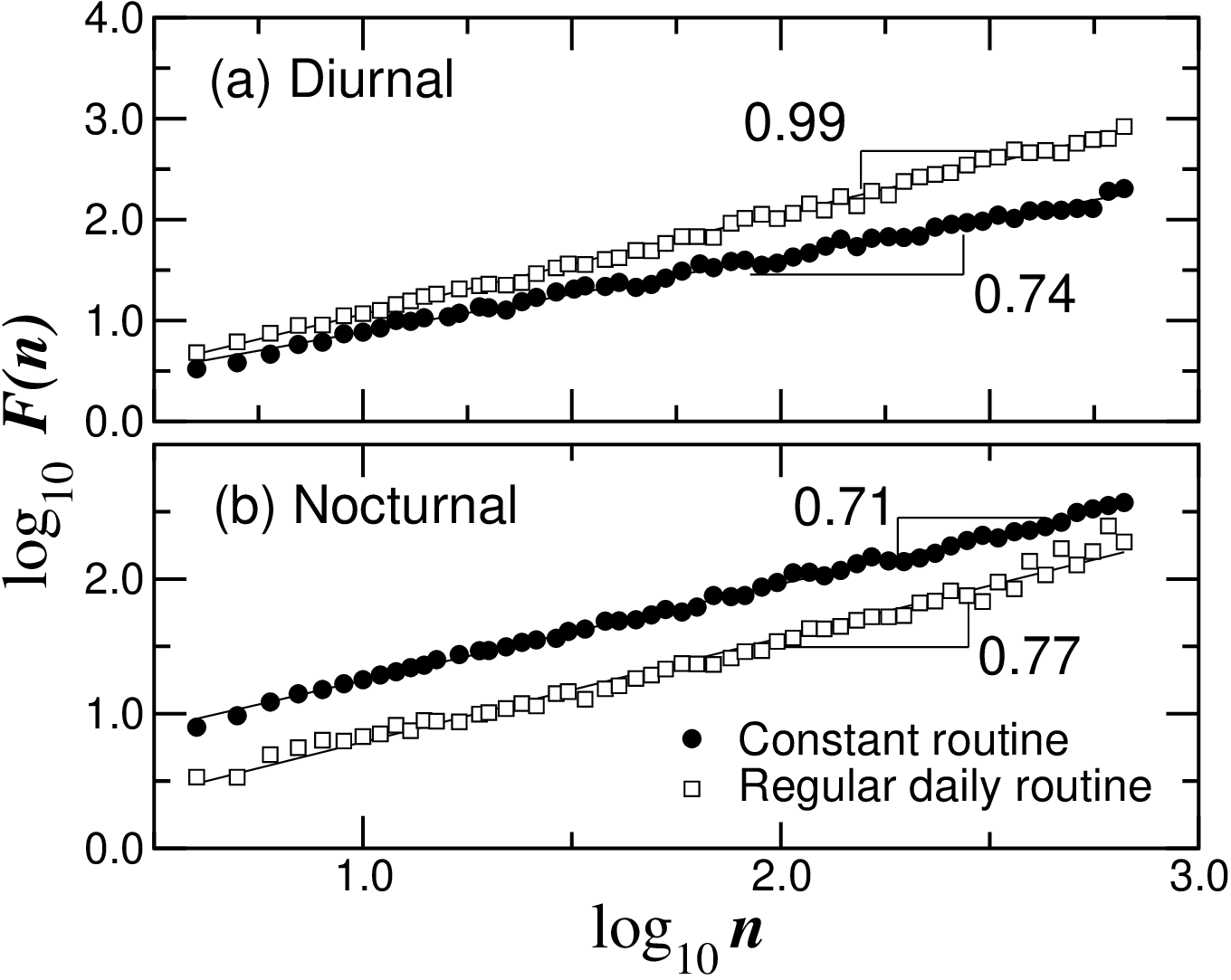} 
 \vspace*{-0.5cm}
 \caption{Long-range time correlations in human voluntary physical
  activity. Double-logarithmic plot of the the DFA function $F(n)$ versus
  time scale $n$ for subject \#2 under the CR protocol and under regular
  daily routine for (a) diurnal and (b) nocturnal periods.  We segmented each
  record into blocks of 2,500 points and performed the DFA analysis on each
  block.  We choose this block length because it provides good time
  resolution while keeping errors in the estimation of the scaling exponents
  small. For all cases, we find close agreement with a linear dependence,
  suggesting a power law increase of $F(n)$.  We tested this assertion by
  considering different time-scale ranges for the power law fit. We found
  that the exponent estimates were nearly independent of the fitting range
  and selected the fitting range $16 \le n \le 200$. For the nocturnal period
  there appears to be no significant difference between the exponent values
  for CR or regular daily routine.  In contrast, there is a marked difference
  between the exponent values for the diurnal period. }
 \label{f.dfa}
\end{figure}

In Fig.~\ref{f.dfa}, we show the results of the DFA analysis for subject \#2
in our database.  We consider separately diurnal and nocturnal periods
because of recent reports indicating altered physiologic control during sleep
\cite{Ivanov,Bunde_sleep}. It is visually apparent that the data for diurnal
and nocturnal periods and for both the CR protocol and the regular daily
routine follow straight lines in the double-logarithmic plots.  This fact
confirms our expectation that Eq.~(\ref{e.dfa}) holds.  Surprisingly, we find
different exponents during diurnal periods for CR and regular daily routine
while we find similar exponent values for the nocturnal periods.  This
finding suggest a difference in the type of voluntary physical activity
\cite{note} during diurnal regular daily routine.

%
%

\begin{figure}[t!]
 \vspace*{0.0cm} 
 \includegraphics*[width=7.6cm]{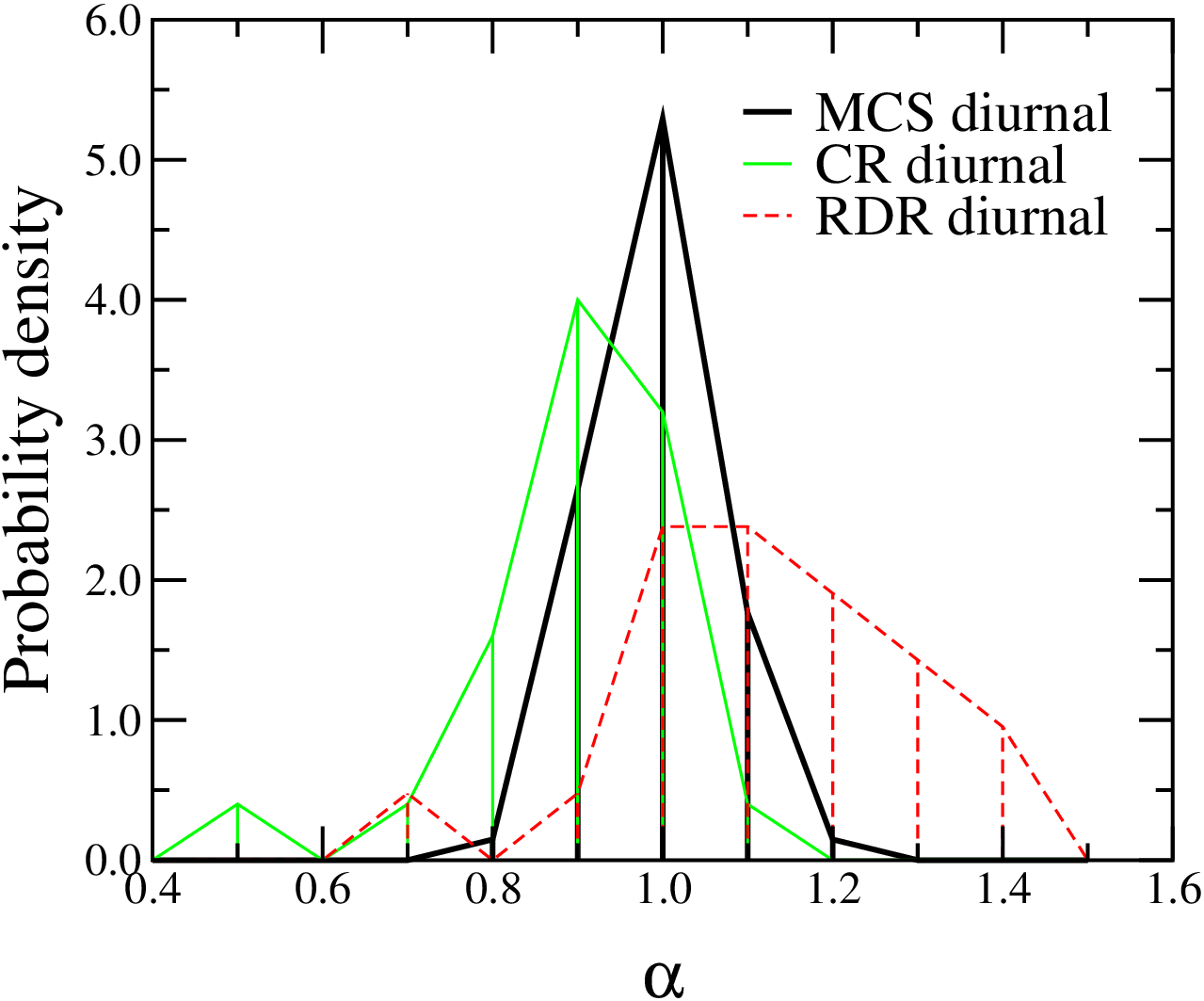} 
 \vspace*{-0.5cm} 
 \caption{ Probability density function of the DFA exponent $\alpha$ of
   diurnal records for the three conditions considered in our study. We find
   that the three distributions are statistically different according to the
   KS test at the 99.99\% confidence level. }
 \label{f.pdf-alpha}
\end{figure}

We test this finding for all other subjects in the database
(Fig.~\ref{f.pdf-alpha} and Table~\ref{t.sig}). Our analysis reveal a value
of $\alpha$ systematically larger for regular daily routine than for CR
during the daytime.  We test this possibility with two statistical test
\cite{nr}: the Student t-test, which tests differences in mean values under
the assumption of Gaussianity of the random variables, and the
Kolmogorov-Smirnov (KS) test, which is a non-parametric test for
distributions.  We find that the difference in the average values of $\alpha$
for the regular daily routine and CR during diurnal periods is significant at
the $p<0.001$ level for both Student's t-test and the KS test.  We also find
that the difference in the average values of $\alpha$ for the regular daily
routine and CR during nocturnal periods is not significant
\cite{note_magnitude}.


To investigate the medical significance of our results for the
characterization of psychiatric and psychosomatic diseases, we next analyze
data collected over a seven day period for four individuals (ages: 34--47 yr)
diagnosed with MCS \cite{mcs-subjects}.  We find that the auto-correlations
of the diurnal physical activity time series are characterized by a DFA
exponent $\alpha= 0.93\pm0.001$ (Table 1 and Fig.~\ref{f.pdf-alpha}) which we
find to be significantly different from the results for healthy subjects at
the $p < 10^{-5}$ level for both the Student's t-test and the KS test.

%
%


A key finding of this study is that voluntary physical activity has a
power-law decaying time correlations whose characteristic index $\alpha$
changes when patterns of physical activity are controlled or disturbed.  For
diurnal periods, we measure a smaller value of $\alpha$ for CR than for
regular daily activity.  The picture emerging from our results raises the
possibility that conditions such as hyperactivity, depression and other
psychiatric and psychosomatic disorders will modify the ``pull'' of daily
routines on physical activity and lead to different values of
$\alpha$. Hence, it is plausible that our finding of fractal correlations in
the activity levels of healthy mature individuals provides a new window into
the study of psychiatric and maturation disorders which is not hindered by
complex (and more falsifiable) procedures such as classifying motion types.
Indeed, we test this possibility for MCS patients and confirm that
correlations in activity are altered by disease.  Further, we have also found
\cite{Kyoko} changes in correlation for patients with a severe fatigue
illness called chronic fatigue syndrome \cite{holmes88,fukuda94}.

Our findings are also of importance because of their implications on the
question of the influence of fractal stimuli, such as physical activity, on
the properties of other rhythms such as heart rate.  Specifically, it has
been recently shown that time series of healthy human interbeat intervals
belong to a special class of complex signals that display multifractal
properties \cite{Ivanov99,multi}.  An explanation is that the
neuroautonomic control mechanisms ---in the presence of even weak external
noise--- {\it endogenously\/} generate multifractal dynamics.

%
\begin{table}[t]
 \caption{ Exponent values for the two protocols and two time periods
  considered in our analysis.  In the table, CR stands for constant routine,
  RDR stands for regular daily routine, and MCS for multiple chemical
  sensitivities, $S$ indicates the number of subjects for which we have
  measurements in each group and $N$ the number of different exponent
  estimates, $\alpha$ and $\sigma_{\alpha}$ are the average and standard
  deviation, respectively, of the DFA exponents for each group. }
 \[
  \begin{tabular}{l|ccccc}
   \textbf{Group}& $S$  & $N$ &$\alpha$ & $\sigma_{\alpha}$ \\
   \hline
   \hline
   CR diurnal    & $7$  & 25 & $0.85\pm0.01$  & $0.03$  \\
   RDR diurnal   & $6$  & 21 & $1.08\pm0.01$  & $0.04$  \\
   MCS diurnal   & $4$  & 68 & $0.93\pm0.001$ & $0.008$ \\
   \hline
   CR nocturnal  & $7$  & 14 & $0.89\pm0.02$  & $0.08$  \\
   RDR nocturnal & $6$  & 12 & $0.95\pm0.04$  & $0.12$  \\
  \end{tabular}
 \]
 \label{t.sig}
\end{table}
%

A recent study showed that the heart rate variability of healthy adults
displays similar multifractal properties for CR and regular daily routine
conditions \cite{Amaral01}.  In contrast, sympathetic or parasympathetic
blockades were found to lead to a significant loss of complexity of HRV
\cite{Amaral01}.  It would thus appear that---in contrast to neuroautonomic
control---voluntary physical activity may not be an important factor in the
generation of the fractal \cite{Meesmann} or multifractal properties of
heartbeat dynamics \cite{Amaral01}.  However, an important assumption for
validating the above conclusion is that voluntary physical activity during
the CR protocol \cite{Constant_rout} {\it is significantly different from
that during regular daily routine}.  Indeed, if during CR conditions there is
only a decrease of the amplitude in the physical activity, then one would
need to show that such a decrease during the CR protocol lowers the amplitude
of the physical activity below a threshold value in order to justify the
endogenous character of the multifractal properties of heartbeat dynamics.
Our finding that the fractal time organization of physical activity during CR
conditions is significantly different from the time organization during
diurnal regular daily routine strengthens the conclusion that voluntary
physical activity is not responsible for the multifractal character of
heartbeat dynamics.

We thank P. Ch. Ivanov, A. L. Goldberger, and H. E. Stanley for stimulating
discussions.  L.A.N.A. thanks the hospitality of the Universidade Federal do
Rio Grande do Norte and support from NIH/NCRR (P41 RR13622) \cite{Physionet}.
D.J.B.S., L.R.S. and L.S.L. acknowledge support from the Brazilian Agencies
CNPq and FINEP.  Y.Y. acknowledges a MECSST Grant-in-Aid for Scientific
Research and support from Japan Science and Technology Corporation.



\begin{thebibliography}{99}
\bibitem{note-cr} This finding confirms the basic
assumption behind the use of the CR protocol---i.e., that the CR protocol
allows the investigation of endogenous rhythms through the minimization of
environmental and behavioral influences.
\bibitem{Tokyo} Each subject gave informed consent to participate in
this study after the test protocol was fully described.  This study
was reviewed and approved by the ethics committee of the Graduate
School of Education of The University of Tokyo.
\bibitem{note} We distinguish between voluntary physical activity, such as
walking, sitting, etc, and involuntary physical activity, such as body
tremor, balance adjustment, etc.
\bibitem{note_magnitude} In this Letter, we do not analyze in detail the
absolute activity level.  The reason is that the use of absolute activity
counts recorded by actigraphs for analysis of human activity has raised a
number of serious concern in the scientific literature (see
Ref.~\protect\cite{na00} for details).  In contrast, the {\it relative\/}
changes measured by the accelerometer, which is the quantity we are concerned
with in this Letter, are not hindered by such methodological concerns
\protect\cite{na00}.
\bibitem{mcs-subjects} Each patient gave informed consent to participate in
this study after the test protocol was fully described.  This study
was reviewed and approved by the ethics committee of The University of
Tokyo Hospital. The method of data collection was the same as that for
healthy subjects.
\end{thebibliography}
\end{document}